\documentclass[conference]{IEEEtran}

\newif\iffull
\usepackage[noend]{algorithmic}
\usepackage{algorithm}
\usepackage[english]{babel}
\usepackage{amssymb}
\usepackage{amsmath}
\usepackage{xcolor}
\usepackage{balance}
\usepackage{cite}
\usepackage{graphicx}
\usepackage{amsthm}
\usepackage{comment}
\usepackage[hidelinks]{hyperref}
\usepackage[a4paper, total={7.15in, 9.8in}]{geometry}

\usepackage{etoolbox}
\makeatletter
\patchcmd{\@makecaption}
  {\scshape}
  {}
  {}
  {}
\makeatletter
\patchcmd{\@makecaption}
  {\\}
  {.\ }
  {}
  {}
\makeatother

\theoremstyle{definition}
\newtheorem{theorem}{Theorem}
\newtheorem{lemma}{Lemma}

\newtheorem{construction}{Construction}

\newtheorem{definition}{Definition}

\newtheorem{example}{Example}

\usepackage{hyperref}
\addto\extrasenglish{

}

\newcommand{\cC}{{\cal C}}

\newcommand{\cE}{{\cal E}}

\newcommand{\cO}{{\cal O}}

\newcommand{\cS}{{\cal S}}

\newcommand{\bfu}{{\boldsymbol u}}
\newcommand{\bfv}{{\boldsymbol v}}

\newcommand{\bfx}{{\boldsymbol x}}
\newcommand{\bfy}{{\boldsymbol y}}

\newcommand{\set}[2]{\left\{#1\;\left|\; #2\right.\right\}}
\newcommand{\abs}[1]{\left|#1\right|}

\title{Optimal Almost-Balanced Sequences}

 \author{%
   \IEEEauthorblockN{\textbf{Daniella~Bar-Lev}\IEEEauthorrefmark{1}, 
                     \textbf{Adir~Kobovich}\IEEEauthorrefmark{2}, 
                     \textbf{Orian~Leitersdorf}\IEEEauthorrefmark{2}, 
                     and \textbf{Eitan~Yaakobi}\IEEEauthorrefmark{1}}
   \IEEEauthorblockA{\IEEEauthorrefmark{1}%
                     Faculty of Computer Science, %\\
                     Technion -- Israel Institute of Technology, 
                     Haifa 3200003, Israel}
\IEEEauthorblockA{\IEEEauthorrefmark{2}%
                     Faculty of Electrical and Computer Engineering, %\\
                     Technion -- Israel Institute of Technology, 
                     Haifa 3200003, Israel}
                     \texttt{\{daniellalev, yaakobi\}@cs.technion.ac.il, \{adir.k, orianl\}@campus.technion.ac.il\vspace{-2ex}}
   \thanks{The research was funded by the European Union (ERC, DNAStorage, 865630). Views and opinions expressed are however those of the authors only and do not necessarily reflect those of the European Union or the European Research Council Executive Agency. Neither the European Union nor the granting authority can be held responsible for them. This work was also supported in part by NSF Grant CCF2212437.}

   }

\IEEEoverridecommandlockouts
\begin{document}

\maketitle

% ---- Abstract ---- %
\begin{abstract}
This paper presents a novel approach to address the constrained coding challenge of generating \emph{almost-balanced sequences}. While strictly balanced sequences have been well studied in the past, the problem of designing efficient algorithms with small redundancy, preferably constant or even a single bit, for almost balanced sequences has remained unsolved. A sequence is \emph{$\boldsymbol{\varepsilon(n)}$-almost balanced} if its Hamming weight is between $\boldsymbol{0.5n\pm \varepsilon(n)}$. It is known that for any algorithm with a constant number of bits, $\boldsymbol{\varepsilon(n)}$ has to be in the order of $\boldsymbol{\Theta(\sqrt{n})}$, with $\boldsymbol{\cO(n)}$ average time complexity. However, prior solutions with a single redundancy bit required $\boldsymbol{\varepsilon(n)}$ to be a linear shift from $\boldsymbol{n/2}$. Employing an iterative method and arithmetic coding, our emphasis lies in constructing almost balanced codes with a single redundancy bit. Notably, our method surpasses previous approaches by achieving the \emph{optimal} balanced order of $\boldsymbol{\Theta(\sqrt{n})}$. Additionally, we extend our method to the non-binary case, considering $\boldsymbol{q}$-ary almost polarity-balanced sequences for even $q$, and almost symbol-balanced for $\boldsymbol{q=4}$. Our work marks the first asymptotically optimal solutions for almost-balanced sequences, for both, binary and non-binary alphabet. 
\end{abstract}

% ---- Introduction ---- %
\section{Introduction}
\label{sec:introduction}
Constrained codes have a long history in information theory, with applications to data storage and transmission. In the broadest setting, raw data in such applications is encoded (in a one-to-one manner) into a set of words $\cS$ over some alphabet $\Sigma$ that satisfy prescribed rules. Some rules are imposed due to physical limitations, such as those dictated by energy compliance or by memory cell wear, and are typically translated into cost constraints. Others are imposed as a preventive measure to keep the storage device in a sufficiently-reliable operation region. A celebrated result in constrained coding theory by Knuth has analyzed strictly balanced binary sequences or sequences with a fixed weight~\cite{knuth}. We consider in this work the \emph{almost-balanced} constraint which generalizes the well-known balanced Knuth codes~\cite{knuth} by requiring that the entire message possess a Hamming weight of \emph{approximately} $n/2$.

One motivating application of this work is DNA storage, where \emph{almost balanced GC content} is necessary~\cite{blawat2016forward}. During the storage phase in DNA strands, media degradation, and in particular breaks, can arise in DNA due to factors that include radiation, humidity, and high temperatures. In \cite{GHPPS15}, the authors proposed to encapsulate the stored DNA in a silica substrate and then to employ custom error-correcting codes to mitigate the effects of these errors. Another approach to dealing with media degradation is to generate strands of DNA that have approximately balanced GC-content, and this approach has been leveraged in several existing works such as \cite{ErlZie17,YazGabMil17,YKGMZM15}.

The construction of efficient balanced codes has been extensively studied; see e.g.~\cite{knuth,TB99,TCB96,IW08,IW10}, and extensions to non-binary balanced codes have been considered in~\cite{tallini1999efficient, MT05,MT06,SW09,WISS13}. Codes that combine the balanced property with certain other constraints, such as run-length limitations, have also been addressed for example in~\cite{IWF11}. However, the problem of almost balanced sequences with Hamming weight between $0.5n\pm \varepsilon(n)$ has received a little attention. Under this framework, the goal is to find the optimal number of redundant bits as a function of $\varepsilon(n)$, where $\varepsilon(n)$ can be a function of $n$, e.g. linear in $n$, $\log n$, or a constant. No less important is the design of such algorithms. 

While Knuth's algorithm is an efficient scheme to strictly balance an arbitrary sequence with $\log n +o(\log n)$ redundancy bits, to design an efficient encoder and decoder with less redundancy or even only a single bit is a non-trivial task. The best known construction that uses a single redundancy bit required $\varepsilon(n)$ to be linear with $n$~\cite{nguyen2020binary}, while a lower bound asserts that the order of $\varepsilon(n)$ has to be $\Omega(\sqrt{n})$. In this work, we close on this gap and present an explicit encoder that uses a single redundancy bit to balance binary sequences for $\varepsilon(n)=\Theta(\sqrt{n})$, with $\cO(n)$ average time complexity.

%\ey{Eitan to complete background of constrained codes. Discuss the paper by Paul. Other papers on balanced codes. Constrained codes for DNA. Locally balanced constraint. Add the paper by the guys from Singapore.}

%In this project we seek to investigate \emph{almost balanced sequences}, which limit the GC content of a string in $\{A,T,C,G\}^n$ to be close to a half, for example between 45\% and 55\%. In its binary analogue, the number of length-$n$ binary vectors where the number of ones is between these two thresholds is given by $\sum_{i=\lceil 0.45n\rceil}^{\lfloor 0.55n \rfloor} {n\choose i}$, and hence the redundancy of this set is calculated to be $n-\log (\sum_{i=\lceil 0.45n\rceil}^{\lfloor 0.55n \rfloor} {n\choose i})<1$, for $n\geq 35$. This bound implies that with at most a single bit of redundancy it is possible to convert an arbitrary sequence to one where the ratio of ones is between 45\% and 55\%. 

The rest of this paper is organized as follows. In \autoref{sec:background} we introduce the definitions that will be utilized throughout the paper and present the arithmetic coding method. Our construction for binary almost-balanced sequences is presented in \autoref{sec:binary} and generalizations for almost polarity-balanced and almost symbol-balanced for non-binary alphabet are presented in \autoref{sec:non-binary}. Section~\ref{sec:conclusion} concludes this paper.

% ---- Background ---- %
\section{Definitions, Related Works, and Arithmetic Coding}
\vspace{-1.5ex}
\label{sec:background}
\subsection{Definitions}\label{subsec:Definitions}
Let $\Sigma_q=\{0,1,\ldots,q-1\}$ be the alphabet of size $q$ and let $\Sigma_q^n$ be the set of all sequences of length $n$ over $\Sigma_q$. The Hamming weight of a sequence $\bfx\in\Sigma_q^n$, denoted by $w(\bfx)$, is the number of non-zero symbols in $\bfx$. The concatenation of two sequences $\bfx$ and $\bfy$ is denoted by $\bfx\circ\bfy$.

A binary sequence $\bfx\in\Sigma_2^n$ is called \emph{balanced} if the number of zeros and ones is identical, i.e., if ${w(\bfx)=\frac{n}{2}}$. We similarly define an \emph{almost-balanced} binary sequence as follows.
\begin{definition}\label{def: binary almost balanced}
    A sequence $\bfx\in\Sigma_2^n$ is called \emph{$\varepsilon(n)$-almost balanced} if $w(\bfx)\in\left[\frac{n}{2}-\varepsilon(n), \frac{n}{2}+\varepsilon(n)\right]$. 
\end{definition}

 To extend the definition of balanced and almost balanced sequences to non-binary sequences we need the following additional notation. For $\sigma\in\Sigma_q$, let $\#_\sigma(\bfx)$ denote the number of occurrences of the symbol $\sigma$ in $\bfx$. A sequence $\bfx\in\Sigma_q^n$ is called \emph{symbol-balanced} if any symbol $\sigma\in\Sigma_q$ appears in $\bfx$ exactly $\frac{n}{q}$ times. That is, $\#_\sigma(\bfx)=\frac{n}{q}$ for any $\sigma\in\Sigma_q$. When $q$ is even, we say that $\bfx$ is \emph{polarity-balanced} if  $$\sum_{i=0}^{\frac{q}{2}-1}\#_{i}(\bfx) = \sum_{i=\frac{q}{2}}^{q-1}\#_{i}(\bfx) = \frac{n}{2}.$$ 

%This definition 
\autoref{def: binary almost balanced} can be extended to \emph{$\alpha$-almost symbol-balanced} and \emph{$\alpha$-almost polarity-balanced} as follows.
\begin{definition}
 A sequence $\bfx$ is called \emph{$\varepsilon(n)$-almost symbol-balanced} if for any $\sigma\in\Sigma_q$ we have that $$\#_{\sigma}(\bfx)\in\left[\frac{n}{q}-\varepsilon(n),\frac{n}{q}+\varepsilon(n)\right].$$
 \end{definition}
\begin{definition}
    A sequence $\bfx$ is called \emph{$\varepsilon(n)$-almost polarity-balanced} if$$\abs{\sum_{i=0}^{\frac{q}{2}-1}\#_{i}(\bfx) - \sum_{i=\frac{q}{2}}^{q-1}\#_{i}(\bfx)} \leq 2\cdot\varepsilon(n).$$
\end{definition}
\subsection{Related Work}\label{subsec:Previous Work}
In this work we extend our previous work~\cite{ConstrainedPeriodicity} focusing on eliminating windows with small periods. This work utilizes a graph-based reduction technique to establish efficiency and convergence of the construction. Inspired by this, the current study proposes an iterative method for encoding sequences into almost-balanced ones without requiring monotonic progress between the algorithm's steps. In a parallel effort~\cite{Kobovich2024universal, bar2023universal}, the technique is extended to address diverse constraints. A universal approach is presented and a general methodology for combining constraints is detailed, showcasing the versatility and comprehensive nature of the encoding framework.

\subsection{Arithmetic Coding}\label{subsec:Arithmetic Coding}
\emph{Arithmetic coding}~\cite{rissanen1979arithmetic} serves as a data compression method wherein the encoding process transforms an input sequence into a new sequence, representing a fractional value in the interval $[0,1)$. Each iteration processes a single symbol from the input, dividing the current interval and designating one of the resulting partitions as the new interval. Consequently, the algorithm progressively operates on smaller intervals, and the output exists within each of these nested intervals. %\ey{is this the descriptio for the binary or arbitrary case?}

One of the main components in our suggested construction is an encoder and decoder pair which are based on  binary arithmetic coding. For the completeness of the results in the paper, the key concepts of binary arithmetic coding, which will be used throughout this paper, are described next.\footnote{This description highlights the details necessary for our derivations and it can be considered a simplification of the standard arithmetic coding technique.}

Let $p\in(0,1)$ and let $n$ be an integer. The encoding of a sequence $\bfx$ of length $n$ is done by mapping $\bfx$ into a unique interval $I_\bfx\subseteq [0,1)$ as follows.  
\begin{enumerate}
    \item Initialize $I\leftarrow[0,1)$.
    \item For $i=1,2,\ldots,n$:
    \begin{enumerate}
        \item Split the interval $I$ into two sub-intervals, $I_L$ and $I_R$, of sizes $|I_L|=p\cdot|I|$ and $|I_R| = (1-p)\cdot |I|$.
        \item If $x_i=0$, $I\leftarrow I_L$
        \item Else, $I\leftarrow I_R$
    \end{enumerate}
    \item $I_\bfx=I$.
\end{enumerate}
Finally, the encoding of $\bfx$ is the binary representation\footnote{Here we consider the binary representation of a fraction in $[0,1)$ as the binary vector representing the corresponding sum of negative powers of two (similar to the representation of a positive integer, but with negative powers). For example $0.25$ is represented by $01$ and $0.75$ is represented by $11$.} of the shortest (fewest number of bits) fraction in the interval $I_\bfx$. 

\begin{figure}[t]
    \centering
    \includegraphics[width=0.98\linewidth]{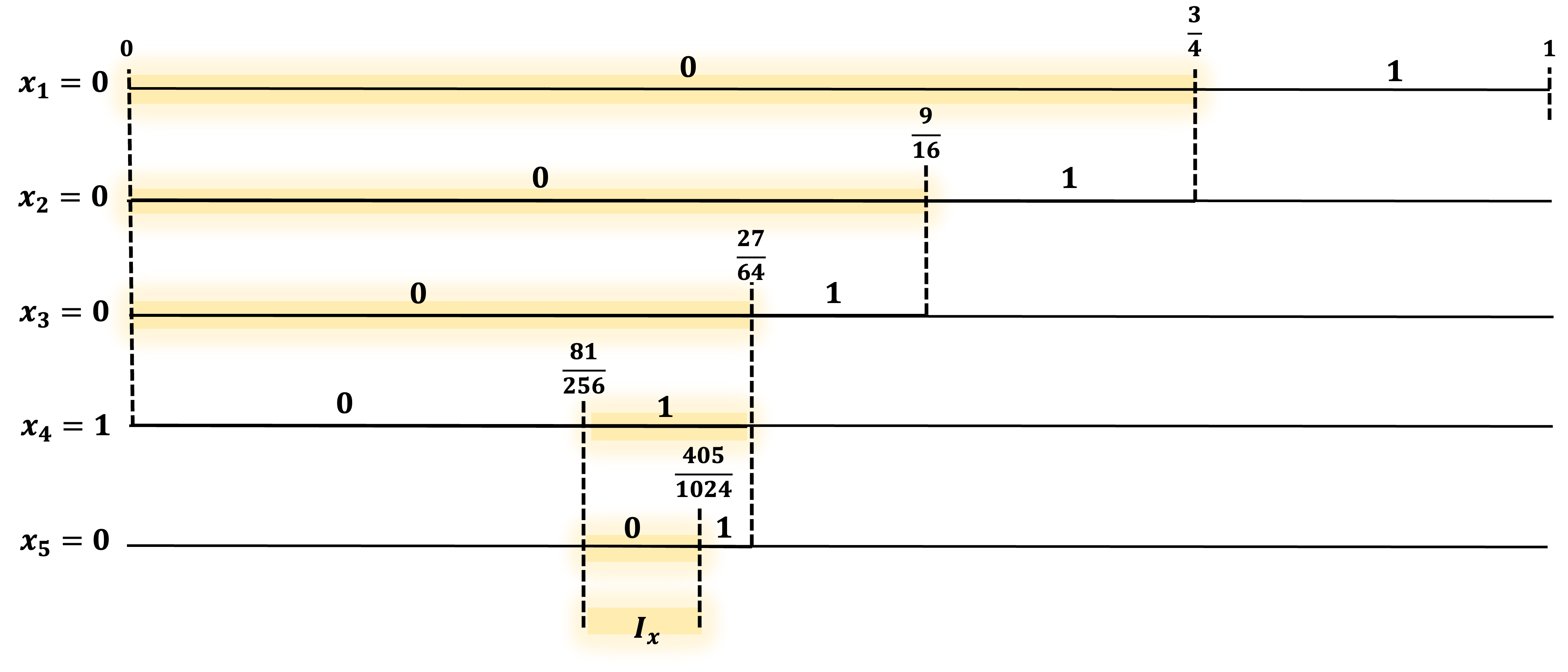}
    \vspace{-2ex}
    \caption{Mapping of $\bfx=00010$ into an interval $I_\bfx$ for $p=\frac{3}{4}$ (and $n=5$).}
    \label{fig:arithmetic_coding}
    \vspace{-1ex}
\end{figure}

After mapping the input to an interval, the output of the algorithm is the binary representation of a fraction within the interval with a minimal number of bits in its binary representation. Given $p\in(0,1)$ and an integer $n>0$ we denote the corresponding encoder and decoder pair by $f_p^{(\textbf{ac})}:\Sigma_2^n\to\Sigma_2^*$ and $g_p^{(\textbf{ac})}:\Sigma_2^*\to\Sigma_2^n$.

\begin{example}
\autoref{fig:arithmetic_coding} presents the described steps for the mapping of $\bfx=00010$ into an interval $I_\bfx$ for $p=\frac{3}{4}$ (and $n=5$). In this example, $I_\bfx = \left[\frac{81}{256},\frac{405}{1024}\right),$
and the fraction with the minimal representation within this interval is $0.375 = 2^{-2}+2^{-3}$, resulting with the output $011$. 
\end{example}

We note that the mapping of a sequence $\bfx\in\Sigma_2^n$ into an interval $I_\bfx$ requires splitting the interval $n$ times, where each iteration computes the new interval edges in $\cO(1)$ time. Hence, the worse-case time complexity of $f_p^{(\textbf{ac})}$ and $g_p^{(\textbf{ac})}$ is $\Theta(n)$.

% ----Binary ---- %
\section{Binary almost balanced sequences}\label{sec:binary}
In this section, we discuss the case of binary $\varepsilon(n)$-almost balanced sequences.  We first explicitly define the constraint as follows,
\begin{align*}
\mathcal{C}(n,\varepsilon(n)) =\set{\bfx \in \Sigma_2^n}{\hspace{-0.5ex}w(\bfx) \in \hspace{-0.5ex}\left[\frac{n}{2}\hspace{-0.3ex}-\hspace{-0.3ex}\varepsilon(n),\frac{n}{2}\hspace{-0.3ex}+\hspace{-0.3ex}\varepsilon(n)\right]}.
\end{align*}

In the next lemma, we show that there exists a single-redundancy-bit construction for $\cC(n,\varepsilon(n))$ only if $\varepsilon(n)=\Omega(\sqrt{n})$. More precisely, for $\varepsilon(n) = \alpha\sqrt{n}$, we give lower and upper bounds on the minimal value of $\alpha$ for which there
exists a single-redundancy-bit construction for $\mathcal{C}(n,\alpha\sqrt{n})$ for $n$ large enough. More formally, for every $\alpha>0$ we define $F(n, \alpha)\triangleq \frac{|\mathcal{C}(n,\alpha\sqrt{n})|}{2^n}$. Thus, our goal is to find the minimum~$\alpha$ for which there exists $n'$ such that for any $n\ge n'$ we have that $F(n, \alpha)\geq 1/2$. 
\begin{lemma}\label{lem: lower bound binary}
%Let $c=\liminf\{\alpha | F(\alpha)\geq 1/2\}$. Then, it holds that $0.335<c\le 0.34$. That is, for $n$ large enough there exists a single-redundancy-bit construction for $\mathcal{C}(n,\alpha\sqrt{n})$ when $\alpha \geq c> 0.335$.
%\begin{lemma}
There exists a constant $c$ such that if $\alpha\ge c$ and~$n$ is large enough, then there exists a single redundancy bit construction for $\mathcal{C}(n,\alpha\sqrt{n})$. Otherwise, if $\alpha< c$ then there is no such a construction. Moreover, it holds that $0.335<c\le 0.34$.
\end{lemma}
%\begin{lemma}
% For $n$ large enough, there exists a constant $c$ such that if $\alpha\ge c$ then there exists a single redundancy bit construction for $\mathcal{C}(n,\alpha\sqrt{n})$. Otherwise, if $\alpha< c$ then there is no such a construction. Moreover, it holds that $0.335<c\le 0.34$.
%\end{lemma}
\begin{proof}
This result can be seen from the fact that the binomial distribution approaches the normal distribution as $n \rightarrow \infty$, with $\mu=n/2$ and $\sigma=\sqrt{n}/2$. Considering the Z-score table~\cite{kreyszig2018advanced} we know that 
 at least half of the space is thus contained in $$[\mu-0.68\sigma, \mu+0.68\sigma] = [n/2-0.34\sqrt{n}, n/2+0.34\sqrt{n}].$$ On the other hand, the interval $$[\mu-0.67\sigma, \mu+0.67\sigma] = [n/2-0.335\sqrt{n}, n/2+0.335\sqrt{n}]$$ contains strictly less than half of the space.\end{proof}
Nonetheless, the problem of finding such a construction has remained unsolved as the state-of-the-art construction~\cite{nguyen2020binary} with a single redundancy bit only tackles a linear-almost-balanced version of $[np_1, np_2]$ for $p_1 < 1/2 < p_2$. It should be noted that the construction in \cite{nguyen2020binary} utilizes the existence of mappings without explicitly determining them. Next, we demonstrate an efficient construction with a single redundancy bit for $\alpha> \sqrt{\ln(2)}\approx 0.8325$, inspired by the approach taken in~\cite{ConstrainedPeriodicity} and based on the arithmetic coding~\cite{rissanen1979arithmetic} technique.
To this end, we also define the two following auxiliary constraints,
\begin{align*}
\mathcal{C}_{L}(n,\alpha\sqrt{n}) = \set{\bfx \in \Sigma_2^n}{w(\bfx)  \leq \frac{n}{2}+\alpha\sqrt{n}},
\end{align*}
\begin{align*}
\mathcal{C}_{H}(n,\alpha\sqrt{n}) = \set{\bfx \in \Sigma_2^n}{w(\bfx) \geq \frac{n}{2}-\alpha\sqrt{n}}.
\end{align*}
Notice that $\mathcal{C}(n,\alpha\sqrt{n}) = \mathcal{C}_{L}(n,\alpha\sqrt{n}) \cap \mathcal{C}_{H}(n,\alpha\sqrt{n})$. We now propose the overall construction as follows,

\begin{construction}[Binary almost-balanced]
\label{const:ab single bit}
Let $\alpha>\sqrt{\ln(2)}$, let $n$ be a sufficiently large\footnote{While the size of $n$ depends on $\alpha$, the size of $n$ for which the construction work is not too large. For example, for $\alpha=1$ and $\alpha=0.835$ we only need $n>4$ and $n>6$, respectively} integer, and let $\bfx\in\{0,1\}^{n-1}$. Our construction is composed of the following two instances of the arithmetic coding described in \autoref{subsec:Arithmetic Coding}:
\begin{itemize}
    \item Binary arithmetic coding with ${p_L={1}/{2}+{\alpha}/{\sqrt{n}}+{1}/{n}}$ and a pair of encoder and decoder functions ${f^{(\textbf{ac})}_{p_L}:\Sigma_2^{n}\to\Sigma_2^*}$, ${g^{(\textbf{ac})}_{p_L}:\Sigma_2^{*}\to\Sigma_2^n.}$ 
    \item Binary arithmetic coding with ${p_H={1}/{2}-{\alpha}/{\sqrt{n}}-{1}/{n}}$ and a pair of encoder and decoder functions ${f^{(\textbf{ac})}_{p_H}:\Sigma_2^{n}\to\Sigma_2^*}$, ${g^{(\textbf{ac})}_{p_H}:\Sigma_2^{*}\to\Sigma_2^n.}$
\end{itemize}
For simplicity, we assume that the output length of $f^{(\textbf{ac})}_{p_L}, f^{(\textbf{ac})}_{p_H}$ is at least $n-2$ (otherwise we can pad the output with zeros and we will show that it will be exactly $n-2$).  
Then, Algorithms~\ref{alg:AB Encoder} and~\ref{alg:AB Decoder} construct an efficient construction with a single redundancy bit and $\cO(T(n))$ average time complexity for $T(n)$ the maximum time complexity amongst $f^{(\textbf{ac})}_{p_L},g^{(\textbf{ac})}_{p_L}, f^{(\textbf{ac})}_{p_H}, g^{(\textbf{ac})}_{p_H}$. That is, the average time complexity is~$\cO(n)$. 
\label{const:universal}
\end{construction}

\begin{algorithm}[t]
    \centering
    \begin{algorithmic}[1]
    \renewcommand{\algorithmicrequire}{\textbf{Input:}}
    \renewcommand{\algorithmicensure}{\textbf{Output:}}
    \REQUIRE $\bfx \in \Sigma_2^{n-1}$.
    \ENSURE $\bfy \in \mathcal{C}(n,\alpha\sqrt{n})$. 
    \STATE $\bfy \gets \bfx\circ 0$.
    
    \WHILE{$\bfy\notin \mathcal{C}(n,\alpha\sqrt{n})$}
        \IF{$\bfy\in \cC_L(n,\alpha\sqrt{n})$}
            \STATE $\bfy \gets f_{p_L}^{(\textbf{ac})}(\bfy)\circ  11$.
        \ELSE
            \STATE $\bfy \gets f_{p_H}^{(\textbf{ac})}(\bfy)\circ 01$.
        \ENDIF
    \ENDWHILE
    
    \RETURN $\bfy$.
    
    \end{algorithmic}
    \caption{Almost-Balanced Encoder $E$}
    \label{alg:AB Encoder}
\end{algorithm}

\begin{algorithm}[t]
    \centering
    \begin{algorithmic}[1]
    \renewcommand{\algorithmicrequire}{\textbf{Input:}}
    \renewcommand{\algorithmicensure}{\textbf{Output:}}
    \REQUIRE $\bfy \in \mathcal{C}(n,\alpha\sqrt{n})$ such that $E(\bfx) = \bfy$ for $\bfx \in \Sigma_2^{n-1}$.
    \ENSURE $\bfx \in \Sigma_2^{n-1}$.
    
    \WHILE{$y_n\ne 0$}
    
    \IF{$y_{n-1}=1$}
            \STATE $\bfy \gets g_{p_L}^{(\textbf{ac})}(\bfy_{[1:n-2]})$.
            \ENDIF
    \IF{$y_{n-1}=0$}
            \STATE $\bfy \gets g_{p_H}^{(\textbf{ac})}(\bfy_{[1:n-2]})$.
        \ENDIF
    
    \ENDWHILE
    
    \RETURN $\bfy_{[1:n-1]}$.
    
    \end{algorithmic}
    \caption{Almost-Balanced Decoder $D$}
    \label{alg:AB Decoder}
\end{algorithm}

The correctness of this construction is stated in the next theorem.
\begin{theorem}
    \autoref{const:ab single bit} is an efficient construction with a single redundancy bit for $\mathcal{C}(n,\alpha\sqrt{n})$ when $\alpha > \sqrt{\ln(2)}$ and $n$ is large enough.
\end{theorem}

The proof of the theorem follows immediately from the following three lemmas. 

\begin{lemma}\label{lem: alg 1 coverges}
    Algorithm~\ref{alg:AB Encoder} stops with an output $\bfy\in\mathcal{C}(n,\alpha\sqrt{n})$. 
\end{lemma}

\begin{proof}
First, let us prove that in each iteration of Algorithm~\ref{alg:AB Encoder}, the length of $\bfy$, is exactly $n$. 
Clearly, before entering the while loop for the first time, the length of $\bfy$ is $n$. Within the while loop, $\bfy$ is modified to be the concatenation of either $f_{p_L}^{(\textbf{ac})}(\bfy)$ or $f_{p_H}^{(\textbf{ac})}(\bfy)$ with two additional bits. Hence, we need to show that the output of $f_{p_L}^{(\textbf{ac})}(\bfy)$ or $f_{p_H}^{(\textbf{ac})}(\bfy)$, respectively, is of length $n-2$. Recall that by our assumption, the output length of $f_{p_L}^{(\textbf{ac})}, f_{p_H}^{(\textbf{ac})}$ is always at least $n-2$. Hence, it is sufficient to show that the length cannot be greater than $n-2$. 

If $\bfy$ is modified in Step $4$ then $\bfy\notin \cC(n,\alpha\sqrt{n})$ and ${\bfy\in \cC_L(n,\alpha\sqrt{n})}$. That is, $w(\bfy)< \frac{n}{2}-\alpha\sqrt{n}$, and the length of the interval that corresponds to $\bfy$ by the mapping  $f_{p_L}^{(\textbf{ac})}$ is 
\begin{equation*}
|I_\bfy|= \left(\frac{1}{2} - \frac{\alpha}{\sqrt{n}}-\frac{1}{n}\right)^{w(\bfy)} \cdot \left(\frac{1}{2} + \frac{\alpha}{\sqrt{n}}+\frac{1}{n}\right)^{n - w(\bfy)}.
\end{equation*}
Since this value is minimized when $w(\bfy)$ is maximized, we have that any such $\bfy$ is mapped to an interval of length 
\begin{equation*}
|I_\bfy|\ge \hspace{-0.5ex}\left(\frac{1}{2} \hspace{-0.5ex}- \hspace{-0.5ex}\frac{\alpha}{\sqrt{n}}\hspace{-0.5ex}-\hspace{-0.5ex}\frac{1}{n}\right)^{\frac{n}{2}\hspace{-0.5ex}-\hspace{-0.5ex}\alpha\sqrt{n}} \cdot \left(\frac{1}{2}\hspace{-0.5ex} +\hspace{-0.5ex} \frac{\alpha}{\sqrt{n}}\hspace{-0.5ex}+\hspace{-0.5ex}\frac{1}{n}\right)^{\frac{n}{2}+\alpha\sqrt{n}}.
\end{equation*}
Moreover, since 
\vspace{-0.5ex}
$$\lim_{n\to\infty}\hspace{-0.5ex} 2^{n-2}\cdot \hspace{-0.5ex}\left(\frac{1}{2} \hspace{-0.5ex}- \hspace{-0.5ex}\frac{\alpha}{\sqrt{n}}\hspace{-0.5ex}-\hspace{-0.5ex}\frac{1}{n}\right)^{\frac{n}{2}\hspace{-0.5ex}-\hspace{-0.5ex}\alpha\sqrt{n}} \cdot \left(\frac{1}{2}\hspace{-0.5ex} +\hspace{-0.5ex} \frac{\alpha}{\sqrt{n}}\hspace{-0.5ex}+\hspace{-0.5ex}\frac{1}{n}\right)^{\frac{n}{2}+\alpha\sqrt{n}} \hspace{-1.5ex}= \frac{e^{2\alpha^2}}{4},$$
it can be verified that for any $\alpha> \sqrt{\ln(2)}$ we have that ${|I_\bfy|\ge 1/2^{n-2}}$ for $n$ large enough.
This implies that it is possible to enumerate the interval with exactly $n-2$ symbols.  

By the definition of Algorithm~\ref{alg:AB Encoder}, if the algorithm ends, it stops with a valid output. Hence, it is left to show that the algorithm converges. Similarly to the approach presented in~\cite{ConstrainedPeriodicity}, the convergence follows from a reduction to an acyclic graph walk, and it is given here for  completeness.

Let $G=(V,\cE)$ be a directed graph such that $V =\Sigma_2^n$ is the set of nodes and the set of edges $\cE\subseteq V\times V$ is defined as follows. From any $\bfu\notin \cC(n,\alpha\sqrt{n})$ there is a single outgoing edge to the node $\bfv\in V$, where $\bfv= f_{p_L}^{(\textbf{ac})}(\bfu)\circ 11$ if $\bfy\in\cC_L(n,\alpha\sqrt{n})$, and otherwise $\bfv = f_{p_H}^{(\textbf{ac})}(\bfu)\circ 01$. That is, there is an edge from node $\bfu$ to node $\bfv$ if $\bfv$ is the unique sequence such that the operations inside the while loop of Algorithm~\ref{alg:AB Encoder} will modify $\bfy=\bfu$ to $\bfy=\bfv$. 
Note that the mappings $f_{p_L}^{(\textbf{ac})}$ and $f_{p_ H}^{(\textbf{ac})}$ are invertible functions and hence the in-degree of all the nodes in $V$ is at most one. Moreover, by the definition of $\cE$, any node $\bfv\in V$ for which $v_n=1$, satisfies $d_{in}(\bfv)=0$. 

Assume by contradiction that the encoder does not converge for an input $\bfx\in\Sigma_2^{n-1}$ and let $$\bfx\circ 0 = \bfy^{(0)},\bfy^{(1)}, \bfy^{(2)}, \ldots$$ be the list of nodes that correspond to the values of $\bfy$ before each iteration of the while loop in Algorithm~\ref{alg:AB Encoder}. Since $|V|$ is finite, the path $\bfy^{(0)}\to\bfy^{(1)}\to\bfy^{(2)}\to\cdots$ contains a cycle, i.e., there is an index $i$ such that 
$$\bfy^{(i)}\to \bfy^{(i+1)}\to\cdots\to\bfy^{(j-1)}\to\bfy^{(j)}=\bfy^{(i)}.$$ For a node $\bfv\in V$, let  $d_{in}(\bfv)$ be the in-degree of $\bfv$. Since $y^{(0)}_n=0$, we have that $d_{in}(\bfy^0)=0$ and hence $\bfy^0$ is not on the cycle. Hence, there exists an index $i'$ such that $d_{in}(\bfy^{(i')})=2$, which is a contradiction.
\end{proof}
\begin{example}
    Let $n=8, \alpha=1$ and $\bfx = 1000000\in\Sigma_2^7$. Note that $\cC(8,\sqrt{8}) = \{\bfy\in\Sigma_2^8\ | \ 2\le w(\bfy)\le 6\}$. Using the same notations as in the proof of \autoref{lem: alg 1 coverges},  Algorithm~\ref{alg:AB Encoder} begins with $\bfy^{(0)} = \bfx \circ 0 = 1000000\circ 0$. As $\bfy^{(0)}\notin \cC(8,\sqrt{8})$ the algorithm enters the while loop and line 4 is executed ($w(\bfy^{(0)})\le 6$). It can be verified that in this step $\bfy^{(0)}$ is mapped to the interval $[0.97855, 0.99698)$ which can be represented using $111111$. Hence $\bfy^{(1)} = 111111\circ 11$. In the second iteration, line $6$ is executed and $\bfy^{(2)} = 100000\circ 01\in \cC(8,\sqrt{8})$. 
\end{example}

\begin{lemma}\label{lem: correct decoder}
    For any $\bfx\in\Sigma_2^{n-1}$, the decoder $D$ from Algorithm~\ref{alg:AB Decoder} satisfies  $D\left(E(\bfx)\right)=\bfx$. 
\end{lemma}

\begin{lemma}\label{lem: time complexity}
    The average number of iterations in the while loop of $E,D$ is at most $\abs{\Sigma} = O(1)$. 
\end{lemma}

The proofs of \autoref{lem: correct decoder} and \autoref{lem: time complexity} follow from the reduction of the encoder to a graph walk presented in the proof of \autoref{lem: alg 1 coverges}. Since the proofs are very similar to the ones presented in \cite{Kobovich2024universal}, they are omitted.

% ---- Non-binary ---- %
\section{Extensions to Non-binary Alphabets}\label{sec:non-binary}
In this section we discuss the extension of \autoref{const:ab single bit} to the non-binary case. 

\subsection{Almost Polarity-Balanced}
\autoref{const:ab single bit} can be modified to obtain almost polarity-balanced sequences for any alphabet of even size. First, we formally define the constraint 

\vspace{-2ex}
\begin{small}
\begin{align*}
\cC_q^{(\textbf{pb})}(n,\varepsilon(n)) \hspace{-0.5ex}& = \hspace{-0.5ex}\set{\hspace{-0.5ex}\bfx\in\Sigma_q^n}{\sum_{i=0}^{\frac{q}{2}-1}\#_{i}(\bfx)  \in \hspace{-0.5ex}\left[\frac{n}{2}-\varepsilon(n), \frac{n}{2}+\varepsilon(n)\right]\hspace{-0.5ex}}.
\end{align*}
\end{small}
Similarly to the binary case, we are interested in single redundancy symbol codes and in the next lemma we show that $\varepsilon(n)=\Omega(\sqrt{n})$. To this end, we define ${F^{(\textbf{pb})}(n, \alpha)\triangleq 
|\cC_q^{(\textbf{pb})}(n,\alpha\sqrt{n})|/q^n}$.

\begin{lemma}
Let $c=\liminf\{\alpha | F^{(\textbf{pb})}(n, \alpha)\geq 1/q\}$. Then, it holds that $c\le 0.34$. 
\end{lemma}
\begin{IEEEproof}
    It holds that
    \begin{align*}
    F^{(\textbf{pb})}(n, \alpha) & = 
    \frac{|\cC_q^{(\textbf{pb})}(n,\alpha\sqrt{n})|}{q^n} 
     =     \frac{|\cC(n,\alpha\sqrt{n})|\cdot(q/2)^n}{q^n} \\
    & =    \frac{|\cC(n,\alpha\sqrt{n})|}{2^n} 
    =  F(n, \alpha).
    \end{align*}
     The latter together with \autoref{lem: lower bound binary} completes the proof. 
\end{IEEEproof}
While the bound in the previous lemma is not tight, knowing that $F^{(\textbf{pb})}(n, \alpha)=F(n, \alpha)$, we can derive tighter upper and lower bounds using the same techniques as in the proof of \autoref{lem: lower bound binary}. Table I summarizes the upper and lower bounds on~$c$ for small alphabet size, that can be derived by this manner. 

\begin{table}[h]
\centering
\caption{Bounds on $c=\liminf\{\alpha | F^{(\textbf{pb})}(n, \alpha)\geq 1/q\}$.}
%\begin{minipage}{304pt}
\label{table bounds}%
\begin{tabular}{c|cc}
\hline
 $q$           & Lower bound  & Upper bound  \\
\hline
2 & 0.335 & 0.34 \\
3 & 0.215 & 0.22 \\
4 & 0.155 & 0.16 \\
5 & 0.125 & 0.13 \\
6 & 0.105 & 0.11 \\
7 & 0.09  & 0.095
\end{tabular}
\end{table}

Similarly to \autoref{const:ab single bit}, we define $\cC_q^{(\textbf{pb})}(n,\alpha\sqrt{n})$ as the intersection of the following two constraint channels,

\vspace{-2ex}
\begin{small}
\begin{align*}
    \cC_{q,L}^{(\textbf{pb})}(n,\alpha\sqrt{n}) \hspace{-0.5ex}& = \set{\hspace{-0.5ex}\bfx\in\Sigma_q^n}{\sum_{i=0}^{\frac{q}{2}-1}\#_{i}(\bfx)   \le  \frac{n}{2}+\alpha\sqrt{n}\hspace{-0.5ex}},\\
    \cC_{q,H}^{(\textbf{pb})}(n,\alpha\sqrt{n}) \hspace{-0.5ex}& = \set{\hspace{-0.5ex}\bfx\in\Sigma_q^n}{\sum_{i=0}^{\frac{q}{2}-1}\#_{i}(\bfx)   \ge  \frac{n}{2}-\alpha\sqrt{n}\hspace{-0.5ex}}.
\end{align*}
\end{small}

To address non-binary alphabets, we modify our arithmetic coding based mappings as follows. First, in each iteration we partition the current interval into $q$ sub-intervals (instead of two) and if the current symbol is $i$, we continue with the $i$-th sub-interval to the next iteration. For $p\in(0,1)$ we define the size of the first $\frac{q}{2}$ sub-intervals to be $\frac{2p}{q}|I|$, and the size of the last $\frac{q}{2}$ sub-intervals to be $\frac{2(1-p)}{q}|I|$, where $I$ is the current interval. Given $p\in(0,1)$, an integer $n>0$, and an even integer $q\ge 2$, we denote the corresponding encoder and decoder pair by $f_{q,p}^{(\textbf{pb-ac})}:\Sigma_q^n\to\Sigma_q^*$ and $g_{q,p}^{(\textbf{pb-ac})}:\Sigma_q^*\to\Sigma_q^n$.

\begin{construction}[Almost polarity balanced]
\label{const:pb-ab single bit}
For an even integer $q\ge 2$, let $\alpha>\sqrt{\ln(q)}$, let $n$ be a sufficiently large integer, and let $\bfx\in\Sigma_q^{n-1}$. Our construction is composed of the following two instances of the modified arithmetic coding:
\begin{itemize}
    \item $q$-ary arithmetic coding with ${p_L=1/2+\alpha/\sqrt{n}+1/n}$ and a pair of encoder and decoder functions ${f^{(\textbf{pb-ac})}_{q,p_L}:\Sigma_q^{n}\to\Sigma_q^*}$, ${g^{(\textbf{pb-ac})}_{q,p_L}:\Sigma_q^{*}\to\Sigma_q^n.}$ 
    \item $q$-ary arithmetic coding with ${p_H=1/2-\alpha/\sqrt{n}-1/n}$ and a pair of encoder and decoder functions ${f^{(\textbf{pb-ac})}_{q,p_H}:\Sigma_q^{n}\to\Sigma_q^*,}$ ${g^{(\textbf{pb-ac})}_{q,p_H}:\Sigma_q^{*}\to\Sigma_q^n.}$
\end{itemize}
Then Algorithms~\ref{alg:AB Encoder} and~\ref{alg:AB Decoder} (with the corresponding modifications) construct an efficient construction with a single redundancy symbol and $\cO(T(n))$ average time complexity, where $T(n)$ is the maximum complexity amongst $f^{(\textbf{pb-ac})}_{q,p_L},g^{(\textbf{pb-ac})}_{q,p_L}, f^{(\textbf{pb-ac})}_{q,p_H}, g^{(\textbf{pb-ac})}_{q,p_H}$.
\end{construction}

The correctness of \autoref{const:pb-ab single bit} follows from the same arguments that were presented in the proof of \autoref{const:ab single bit}. Hence, it is sufficient to show that 
the length of $\bfy$ before each iteration of the while-loop is $n$. That is, we need to show that
\begin{align*}
\left(\hspace{-0.5ex}\frac{1}{q} \hspace{-0.5ex}- \hspace{-0.5ex}\frac{2\alpha}{q\sqrt{n}}\hspace{-0.5ex}-\hspace{-0.5ex}\frac{2}{qn}\hspace{-0.5ex}\right)^{\hspace{-0.8ex}\frac{n}{2}\hspace{-0.5ex}-\hspace{-0.5ex}\alpha\sqrt{n}} \hspace{-0.8ex} \cdot \hspace{-0.8ex} \left(\hspace{-0.5ex}\frac{1}{q}\hspace{-0.5ex} +\hspace{-0.5ex} \frac{2\alpha}{q\sqrt{n}}\hspace{-0.5ex}+\hspace{-0.5ex}\frac{2}{qn}\hspace{-0.5ex}\right)^{\hspace{-0.8ex}\frac{n}{2}+\alpha\sqrt{n}}\ge \frac{1}{q^{n-2}},
\end{align*}
which holds for any $\alpha>\sqrt{\ln(q)}$. 
We note that for ${q=2}$ \autoref{const:pb-ab single bit} is identical to \autoref{const:ab single bit}. However, for ${q>2}$ we can improve \autoref{const:pb-ab single bit} by noticing that now instead of using two bits it is enough to use one symbol in order to indicate the three options of determining when the decoder stops and whether to decode with $g^{(\textbf{pb-ac})}_{q,p_L}$ or $g^{(\textbf{pb-ac})}_{q,p_H}$. 
%, and these options can be marked by one symbol. \ey{fix this sentence to say better for binary we need two bits and now one symbol is sufficient for $q\geq 3$.} 
Hence, we can change Algorithm~\ref{alg:AB Encoder} such that in step~$4$ we assign $f_{q,p_L}^{(\textbf{pb-ac})}(\bfy)\circ 1$ into $\bfy$ and in step $6$ we assign $f_{q,p_L}^{(\textbf{pb-ac})}(\bfy)\circ 2$ into $\bfy$. Accordingly, we modify step $2$ and step $5$ of Algorithm~\ref{alg:AB Decoder} to check whether $y_n$ equals $1$ or $2$, respectively.  
By doing so, we allow the output of $f_{q,p_L}^{(\textbf{pb-ac})}$ and $f_{q,p_H}^{(\textbf{pb-ac})}$ to be of length $n-1$. Hence, we need
\begin{align*}
\left(\hspace{-0.5ex}\frac{1}{q} \hspace{-0.5ex}- \hspace{-0.5ex}\frac{2\alpha}{q\sqrt{n}}\hspace{-0.5ex}-\hspace{-0.5ex}\frac{2}{qn}\hspace{-0.5ex}\right)^{\hspace{-0.8ex}\frac{n}{2}\hspace{-0.5ex}-\hspace{-0.5ex}\alpha\sqrt{n}} \hspace{-0.8ex} \cdot \hspace{-0.8ex} \left(\hspace{-0.5ex}\frac{1}{q}\hspace{-0.5ex} +\hspace{-0.5ex} \frac{2\alpha}{q\sqrt{n}}\hspace{-0.5ex}+\hspace{-0.5ex}\frac{2}{qn}\hspace{-0.5ex}\right)^{\hspace{-0.8ex}\frac{n}{2}+\alpha\sqrt{n}}\ge \frac{1}{q^{n-1}},
\end{align*}
which holds for $\alpha>\sqrt{\frac{\ln(q)}{2}}$.

A more thorough examination, reveals that the latter construction can be further improved for larger values of $q$. For such values of $q$, we do not need $f_{q,p_L}^{(\textbf{pb-ac})}$ and $f_{q,p_H}^{(\textbf{pb-ac})}$ to compress the input at all. For our purposes, it is sufficient to only restrict the values of the last symbol in the output and use the remaining symbols to distinguish between the states.

\subsection{Almost symbol-balanced}
Lastly, we discuss $\varepsilon(n)$-almost symbol-balanced $q$-ary sequences. For simplicity, we only give a construction for $q=4$ while the generalization to  $q=2^\ell$ for any positive integer $\ell$ is straightforward. Given $n$ and $\varepsilon$ we define the constraint as,

\vspace{-2ex}
\begin{small}
\begin{align*}
\cC_4^{(\textbf{sb})}(n,\varepsilon(n)) \hspace{-0.5ex} = \hspace{-0.5ex}\set{\hspace{-0.5ex}\bfx\in\Sigma_4^n\hspace{-0.5ex}}{\hspace{-0.5ex}\#_\sigma(\bfx)\hspace{-0.5ex}\in\hspace{-0.5ex}\left[\frac{n}{4}\hspace{-0.5ex}-\hspace{-0.5ex}\varepsilon(n),\frac{n}{4}\hspace{-0.5ex}+\hspace{-0.5ex}\varepsilon(n)\right]\hspace{-0.5ex}, \forall\sigma\in\Sigma_4\hspace{-0.5ex}}.
\end{align*}
\end{small}

We start by noting that any $\varepsilon(n)$-almost symbol-balanced  sequence is also a $ \varepsilon(n)$-almost polarity-balanced. Hence, our analysis of almost polarity-balanced codes implies that if a single-redundancy-symbol $\varepsilon(n)$-almost symbol-balanced code exists then $\varepsilon(n) = \Omega(\sqrt{n})$. Therefore, we focus again on the case where $\varepsilon(n)=\alpha\sqrt{n}$.

Our construction is based on defining a subset of $\cC_4^{(\textbf{sb})}(n,\alpha\sqrt{n}))$ as the intersection of the following three $\frac{\alpha\sqrt{n}}{2}\text{-polarity-balanced}$ $4$-ary codes, 

\vspace{-2ex}
\begin{small}
\begin{align*}
\cC_{0,i}^{(\textbf{pb})} \hspace{-0.5ex} & = \hspace{-0.5ex}\set{ \bfx\in\Sigma_4^n}{\hspace{-0.5ex} \#_{0}(\bfx)  +  \#_{i}(\bfx) \hspace{-0.5ex} \in \hspace{-0.5ex}\left[\frac{n}{2} - \frac{\alpha\sqrt{n}}{2}, \frac{n}{2} + \frac{\alpha\sqrt{n}}{2}\right]},%\\
%\cC_{0,2}^{(\textbf{pb})} \hspace{-0.5ex} & = \hspace{-0.5ex}\set{ \bfx\in\Sigma_4^n}{\hspace{-0.5ex} \#_{0}(\bfx)  +  \#_{2}(\bfx) \hspace{-0.5ex} \in \hspace{-0.5ex}\left[\frac{n}{2} - \frac{\alpha\sqrt{n}}{2}, \frac{n}{2} + \frac{\alpha\sqrt{n}}{2}\right]},\\
%\cC_{0,3}^{(\textbf{pb})} \hspace{-0.5ex} & = \hspace{-0.5ex}\set{ \bfx\in\Sigma_4^n}{\hspace{-0.5ex} \#_{0}(\bfx)  +  \#_{3}(\bfx) \hspace{-0.5ex} \in \hspace{-0.5ex}\left[\frac{n}{2} - \frac{\alpha\sqrt{n}}{2}, \frac{n}{2} + \frac{\alpha\sqrt{n}}{2}\right]}.
\end{align*}
for $i\in\{1,2,3\}$.
\end{small}

\begin{lemma}
    It holds that $\cC_{0,1}^{(\textbf{pb})}\cap \cC_{0,2}^{(\textbf{pb})} \cap \cC_{0,3}^{(\textbf{pb})} \subseteq \cC_4^{(\textbf{sb})}(n,\alpha\sqrt{n}).$
\end{lemma}

\begin{construction}[Almost symbol balanced]
\label{const:sb single symbol}
Let ${\alpha>2\sqrt{\ln(4)}}$, let $n$ be a sufficiently large integer, and let $\bfx\in\Sigma_q^{n-1}$. Our construction is composed of three pairs of the modified arithmetic coding that were utilized in \autoref{const:pb-ab single bit}. For each $i\in\{1,2,3\}$ consider $\cC_{0,i}^{(\textbf{pb})}$ and define:
\begin{itemize}
    \item $4$-ary arithmetic coding with $p_L=1/2+\alpha/2\sqrt{n}+1/n$ that associates the first two intervals in each partition with~$0$ and $i$, and a pair of encoder and decoder functions
        $f^{(\textbf{pb-ac})}_{4,p_L, i}:\Sigma_4^n\to\Sigma_4^*$, $g^{(\textbf{pb-ac})}_{4,p_L, i}:\Sigma_4^*\to\Sigma_4^n.$
    \item $4$-ary arithmetic coding with $p_H=1/2-\alpha/2\sqrt{n}-1/n$ that associates the first two intervals in each partition with~$0$ and $i$, and a pair of encoder and decoder functions $f^{(\textbf{pb-ac})}_{4,p_H, i}:\Sigma_4^n\to\Sigma_4^*$, $g^{(\textbf{pb-ac})}_{4,p_H, i}:\Sigma_4^*\to\Sigma_4^n.$
    \item The constraint channels 
     \begin{align*}
     \cC_{L,0,i}^{(\textbf{pb})}&  = \set{\bfx\in\Sigma_4^n}{\#_0(\bfx)+\#_i(\bfx)\le \frac{n}{2} + \frac{\alpha\sqrt{n}}{2}} \\
     \cC_{H,0,i}^{(\textbf{pb})}&  = \set{\bfx\in\Sigma_4^n}{\#_0(\bfx)+\#_i(\bfx)\ge \frac{n}{2} - \frac{\alpha\sqrt{n}}{2}}
     \end{align*}
\end{itemize}
Then Algorithms~\ref{alg:symbol AB Encoder} and~\ref{alg:symbol AB Decoder} construct an efficient construction with a single redundancy symbol and $\cO(T(n))$ average time complexity, where $T(n)$ is the maximum complexity amongst $f^{(\textbf{pb-ac})}_{4,p_L, i}, f^{(\textbf{pb-ac})}_{4,p_H, i}, g^{(\textbf{pb-ac})}_{4,p_L, i}, g^{(\textbf{pb-ac})}_{4,p_H, i}$ for $i\in\{1,2,3\}$. 
\end{construction}

\begin{algorithm}[t]
    \centering
    \begin{algorithmic}[1]
    \renewcommand{\algorithmicrequire}{\textbf{Input:}}
    \renewcommand{\algorithmicensure}{\textbf{Output:}}
    \REQUIRE $\bfx \in \Sigma_4^{n-1}$.
    \ENSURE $\bfy \in \cC_4^{(\textbf{sb})}(n,\alpha\sqrt{n})$. 

    \STATE $\bfy \gets \bfx\circ 0$.
    
    \WHILE{$\bfy\notin \cC_4^{(\textbf{sb})}(n,\alpha\sqrt{n})$}
    \FOR{$i\in\{1,2,3\}$}    
        \IF{$\bfy\notin \cC_{0,i}^{\textbf{(pb)}}$}
            \IF{$\bfy\notin\cC_{L,0,i}^{(\textbf{pb})}$}
            \STATE $\bfy \gets  f^{(\textbf{pb-ac})}_{4,p_L, i}(\bfy)\circ  1\circ i$.
        \ELSE
            \STATE $\bfy \gets  f^{(\textbf{pb-ac})}_{4,p_H, i}(\bfy)\circ  0\circ i$.
            \ENDIF
        \ENDIF
        \STATE \textbf{break}
        \ENDFOR
    \ENDWHILE
    
    \RETURN $\bfy$.
    
    \end{algorithmic}
    \caption{Almost-Balanced $4$-ary Encoder $E_4$}
    \label{alg:symbol AB Encoder}
\end{algorithm}

\begin{algorithm}[t]
    \centering
    \begin{algorithmic}[1]
    \renewcommand{\algorithmicrequire}{\textbf{Input:}}
    \renewcommand{\algorithmicensure}{\textbf{Output:}}
    \REQUIRE $\bfy \in \cC_4^{(\textbf{sb})}(n,\alpha\sqrt{n})$ such that $E_4(\bfx) = \bfy$ for some $\bfx \in \Sigma_4^{n-1}$.
    \ENSURE $\bfx \in \Sigma_4^{n-1}$.
    
    \WHILE{$y_n\ne 0$}
    
    \IF{$y_{n-1}=1$}
            \STATE $\bfy \gets g_{4,p_L,y_n}^{(\textbf{pb-ac})}(\bfy_{[1:n-2]})$.
            \ENDIF
    \IF{$y_{n-1}=0$}
            \STATE $\bfy \gets g_{4,p_H,y_n}^{(\textbf{pb-ac})}(\bfy_{[1:n-2]})$.
        \ENDIF
    
    \ENDWHILE
    
    \RETURN $\bfy_{[1:n-1]}$.
    
    \end{algorithmic}
    \caption{Almost-Balanced $4$-ary Decoder $D_4$}
    \label{alg:symbol AB Decoder}
\end{algorithm}

The correctness of \autoref{const:sb single symbol} follows from arguments similar to the those presented in the proof of \autoref{const:ab single bit}, and the observation that for $i\in\{1,2,3\}$ the output of  $f^{(\textbf{pb-ac})}_{4,p_H, i}$ and  $f^{(\textbf{pb-ac})}_{4,p_H, i}$ is of length $n-2$ for $\alpha>2\sqrt{\ln(4)}$.

\section{Conclusion}
\label{sec:conclusion}

This work studies the problem of encoding almost-balanced sequences using a single redundancy symbol. While our constructions achieve an optimally balanced order of $\Theta(\sqrt{n})$, there persists a multiplicative gap between theoretical bounds on $\varepsilon(n)$ and the values applicable in our constructions. Furthermore, bounding the worst case time complexity of the algorithms is a challenging problem which is left for future work. Experimental results verified that the number of iterations of Algorithm~\ref{alg:AB Encoder} is at most 7 for words of length $n=30$. 

%\section*{Acknowledgments}

\bibliographystyle{IEEEtran}
\bibliography{almost-balanced}

\balance

\end{document}